\documentclass[letterpaper, 10 pt, conference]{ieeeconf}  % Comment this line out
\IEEEoverridecommandlockouts                              % This command is only
\title{\LARGE \bf
Scalable Planning in Multi-Agent MDPs
}

\author{Dinuka Sahabandu$^{\ast}$, Luyao Niu$^{\ast}$, Andrew Clark, and Radha Poovendran% <-this % stops a space
\thanks{D. Sahabandu and R. Poovendran are with the Network Security Lab, Department of Electrical and Computer Engineering, University of Washington, Seattle, WA 98195-2500, USA {\tt\small \{sdinuka,rp3\}@uw.edu}}%
\thanks{L. Niu and A. Clark are with the Department of Electrical and Computer Engineering, Worcester Polytechnic Institute,
        Worcester, MA 01609, USA
        {\tt\small \{lniu,aclark\}@wpi.edu}}%
\thanks{$^{\ast}$Authors contributed equally to this work.}%
}

\usepackage{amsmath}
\usepackage{amssymb}
\usepackage{color}

\usepackage{algorithmicx}
\usepackage{algorithm}
\usepackage{algpseudocode}

\usepackage{graphics}
\usepackage{graphicx}

\usepackage{adjustbox}
\usepackage{makecell}

\newtheorem{definition}{Definition}
\newtheorem{lemma}{Lemma}
\newtheorem{proposition}{Proposition}
\newtheorem{theorem}{Theorem}

\begin{document}

\maketitle
\thispagestyle{empty}
\pagestyle{empty}

%%%%%%%%%%%%%%%%%%%%%%%%%%%%%%%%%%%%%%%%%%%%%%%%%%%%%%%%%%%%%%%%%%%%%%%%%%%%%%%%
\begin{abstract}

Multi-agent Markov Decision Processes (MMDPs) arise in a variety of applications including target tracking, control of multi-robot swarms, and multiplayer games. A key challenge in MMDPs occurs when the state and action spaces grow exponentially in the number of agents, making computation of an optimal policy computationally intractable for medium- to large-scale problems. One property that has been exploited to mitigate this complexity is transition independence, in which each agent’s transition probabilities are independent of the states and actions of other agents. Transition independence enables factorization of the MMDP and computation of local agent policies but does not hold for arbitrary MMDPs. In this paper, we propose an approximate transition dependence property, called $\delta$-transition dependence and develop a metric for quantifying how far an MMDP deviates from transition independence. Our definition of $\delta$-transition dependence recovers transition independence as a special case when $\delta$ is zero. We develop a polynomial time algorithm in the number of agents that achieves a provable bound on the global optimum when the reward functions are monotone increasing and submodular in the agent actions.
% We develop algorithms that run in polynomial time in the number of agents and achieve a provable bound on the global optimum when the reward functions are monotone increasing and submodular in the agent actions. 
We evaluate our approach on two case studies, namely, multi-robot control and multi-agent patrolling example. %In the target-tracking scenario, we find that our approach is within 5\% of the true optimum, while the stochastic security game achieves at least 1/2-optimality even for large numbers of agents. In both cases, our approach provides a \textbf{***} speedup compared to computing the global optimal policy.

\end{abstract}

\section{Introduction}
\label{sec:intro}
A variety of distributed planning and decision-making problems, including multiplayer games, search and rescue, and infrastructure monitoring, can be modeled as Multi-agent Markov Decision Processes (MMDPs). In such processes, the state transitions and rewards are determined by the joint actions of all of the agents. While there is a substantial body of work on computing such optimal joint policies \cite{wang2003reinforcement,littman2001value,lauer2000algorithm}, a key challenge is that the the total number of states and actions grows exponentially in the number of agents. This increases the complexity of  computing an optimal policy, as well as storing and implementing the policy on the agents.

One approach to mitigating this complexity is to identify additional problem structures. One such structure is \emph{transition-independence} (TI) \cite{becker2004solving}. In a TI-MDP, the state transitions of an agent are independent of the states and actions of the other agents. Such MDPs may arise, for example, in multi-robot scenarios where the motion of each robot is independent of the others. TI-MDPs can be approximately solved by factoring into multiple MDPs, one for each agent, and then obtaining a \emph{local policy}, in which each agent’s next action depends only on that agent’s current state. When the TI property holds and the MDP possesses additional structure, such as submodularity, this approach may yield scalable algorithms for computing near-optimal policies \cite{kumar2017decentralized}.

The TI property, however, does not hold for general MMDPs when there is coupling between the agents. Coupling occurs when two agents must cooperate to reach a particular state, or when the actions of agents may interfere with each other. In this case, the existing results providing near-optimality do not hold, and at present there are no scalable algorithms for local policy selection in non-TI-MMDPs.

In this paper, we investigate the problem of computing approximately optimal local policies for non-TI MMDPs. We propose $\delta$-transition dependence, which captures the deviation of the MMDP from transition independence.  We make the following  contributions:
\begin{itemize}
\item We define the $\delta$-transition dependence property, in which the parameter $\delta$ characterizes the maximum change in the probability distribution of any agent due to changes in the states and actions of the other agents.
\item We propose a local search algorithm for computing local policies of the agents, in which each agent computes an optimal policy assuming that the remaining agents follow given, fixed policies.
\item We prove that, when the reward functions are monotone and submodular in the agent actions, the proposed algorithm achieves a provable optimality bound as a function of the dependence parameter $\delta$ and the ergodicity of the MMDP.
%\textbf{***} optimality bound when the reward functions are monotone and submodular in the agent actions, where \textbf{**parameter names}.
\item We evaluate our approach on two numerical case studies, namely, a patrolling example and a multi-robot target tracking scenario. On average, 
our approach achieves $0.95$-optimality in the multi-robot scenario and $0.99$-optimality in the multi-agent patrolling example, while requiring 10-20\% of the runtime of an exact optimal algorithm.
\end{itemize}
The paper is organized as follows. Section \ref{sec:related} presents the related work. Section \ref{sec:background} contains preliminary results. Section \ref{sec:formulation} presents our problem formulation and  algorithm. Section \ref{sec:optimality} contains optimality and complexity analyses. Section \ref{sec:simulation} presents simulation results. Section \ref{sec:conclusion} concludes the paper.

%We then give our problem formulation and proposed algorithm. We present our optimality and complexity analysis. We then present numerical results and conclude the paper.
\section{Related Work}
\label{sec:related}
MDPs have been extensively studied as a framework for multi-agent planning and decision-making~\cite{littman1994markov,parsons2002game}. Most existing works focus on selecting an optimal joint strategy for the agents, which maps each global system state to an action for each agent \cite{wang2003reinforcement,littman2001value,lauer2000algorithm}. These  methods can be shown to converge to a locally optimal policy, in which no agent can improve the overall reward by unilaterally changing its policy. These joint decision-making problems can be viewed as special cases of multi-agent games in which all agents have a shared reward \cite{zhang2019multi}. These approaches, however, suffer from a ``curse of dimensionality,'' in which the state space grows exponentially in the number of agents, and hence do not scale well to large numbers of agents.

Transition-independent MDPs (TI-MDPs) provide  problem structure that can be exploited to speed up the computation~\cite{becker2004solving}. In a TI-MDP, each agent's transitions probabilities are independent of the actions and states of the other agents, allowing the MDP to be factored and approximately solved~\cite{guestrin2003efficient,becker2004decentralized,beynier2005polynomial,gupta2019successor}. Extensions of the TI-MDP approach to POMDPs were presented in~\cite{amato2013decentralized,amato2019modeling}. A greedy algorithm for TI-MDPs with submodular rewards was proposed in~\cite{kumar2017decentralized}. The goal of the present paper is to extend these works to non-TI MDPs by relaxing  transition independence, enabling optimality bounds for a broader class of MDPs. A local policy algorithm was proposed in \cite{qu2019exploiting} that leverages a fast-decaying property that is distinct from the approximate transition independence that we consider.

Our optimality bounds rely on submodularity of the reward functions. Submodularity is a diminishing-returns property of discrete functions that has been studied in a variety of contexts, including offline~\cite{fisher1978analysis}, online~\cite{buchbinder2014online}, adaptive~\cite{golovin2011adaptive}, and streaming~\cite{badanidiyuru2014streaming} submodularity. Submodular properties were leveraged to improve the optimality bounds of multi-agent planning \cite{kumar2017decentralized}, sensor scheduling~\cite{satsangi2015exploiting}, and solving POMDPs~\cite{albright1979structural}. Submodularity for transition-dependent MDPs, however, has not been explored.
\section{Background and Preliminaries}
\label{sec:background}
This section gives  background on perturbations of Markov chains, as well as definition and relevant properties of submodularity. 

\subsection{Perturbations of Markov Chains}
\label{subsec:Markov-chain-background}
A finite-state, discrete-time Markov chain is a stochastic process defined over a finite set $S$, in which the next state is chosen according to a probability distribution $P(s, \cdot)$, where $s \in S$ is the current state. A Markov chain over $S$ is defined by its transition matrix $P$, in which $P(s,s^{\prime})$ represents the probability of a transition from state $s$ to state $s^{\prime}$. The following theorem describes the steady-state behavior of a class of Markov chains.
\begin{theorem}[Ergodic Theorem \cite{schutze2008introduction}]
\label{theorem:ergodic}
Consider a Markov chain with transition matrix $P$. Suppose there exists $T_{0} > 0$ such that $(P^{t})_{ij} > 0$ for all $t > T_0$ and $i,j \in S$.  Then there is a probability distribution $\pi$ over $S$ such that, for any distribution over the initial state, $$\lim_{t \rightarrow \infty}{\frac{\eta(s,t)}{t}} = \pi(s),$$ where $\eta(s,t)$ is the number of times the Markov chain reaches state $s$ in the first $t$ time steps. Moreover, $\pi$ is the unique left eigenvector of $P$ with eigenvalue $1$.
\end{theorem}

A Markov chain satisfying the conditions of Theorem \ref{theorem:ergodic} is \emph{ergodic}. The probability distribution defined in Theorem \ref{theorem:ergodic} is the \emph{stationary distribution} of the chain. Intuitively, a Markov chain is ergodic if the relative frequency of reaching each state is independent of the initial state. The \emph{ergodicity coefficient} of a matrix $P$ is defined by $$\Lambda_{1}(P) = \frac{1}{2}\max_{i,j}{\sum_{k}{|P_{ik}-P_{jk}|}}.$$ 

We next state preliminary results on perturbations of ergodic Markov chains. First, we define the total variation distance between two probability distributions as follows. For two probability distributions $\mu$ and $\nu$ over a finite space $\Omega$, the \emph{total variation distance} is defined by $$||\mu-\nu||_{TV} \triangleq \max_{\Theta \subseteq \Omega}{|\mu(\Theta) - \nu(\Theta)|}.$$ The total variation distance satisfies \cite{levin2017markov} $$||\mu-\nu||_{TV} = \frac{1}{2}\sum_{x \in \Omega}{|\mu(x)-\nu(x)|}.$$ 

%The total variation distance can be bounded via the method of coupling, defined as follows. A \emph{coupling} between two random variables $\mu$ and $\nu$ is a pair of random variables $X$ and $Y$ such that the marginal distributions of $X$ and $Y$ are $\mu$ and $\nu$, respectively. The following result gives a coupling-based approach to bounding the total variation distance.

%\begin{lemma}[\cite{levin2017markov}, Proposition 4.7]
%\label{lemma:coupling}
%The total variation distance between distributions $\mu$ and $\nu$ can be bounded by
%\begin{equation}
 %   \label{eq:TV-coupling}
    %|\mu-\nu||_{TV} = \inf{\{Pr(X\neq Y): (X,Y) \mbox{ is a coupling of } \mu \mbox{ and } %\nu\}}
%\end{equation}
%\end{lemma}
    %We next give needed background on perturbations of Markov chains. 
    Let $P$ and $P^{\prime}$ denote the transition matrices of two ergodic Markov chains on the same state space with stationary distributions $\mu$ and $\nu$, and define $\Delta = P-P^{\prime}$. The $1$-norm of the matrix $\Delta$ is defined by $$||\Delta||_{1} = \max_{i}{\left\{\sum_{j}{|\Delta_{ij}|}\right\}},$$ where $\Delta_{ij}$ is the $(i,j)$-th entry of $\Delta$. The group inverse of $P$, denoted $P^{\#}$, is the unique square matrix satisfying $$PP^{\#}P = P, \ P^{\#}PP^{\#} = P^{\#}, \ P^{\#}P = PP^{\#}.$$ Let $Z = I-P$, where $I$ denotes the identity. It is known \cite{meyer1975role} that $Z^{\#} = (I-P+\mu\mathbf{1}^{T})^{-1}-\mu\mathbf{1}^{T}$, where $\mathbf{1}$ denotes the vector with all $1$'s. 
    
    The following result gives a bound on the distance between $\mu$ and $\nu$ as a function of the perturbation $\Delta$.
    
    \begin{lemma}[\cite{seneta1991sensitivity}]
    \label{lemma:perturbation}
    The total variation distance between the stationary distributions $\mu$ and $\nu$ of Markov chains with transition matrices $P$ and $P^{\prime}$, respectively, satisfies $$||\mu-\nu||_{TV} \leq \frac{1}{2}\Lambda_{1}(Z^{\#})||P-P^{\prime}||_{1},$$ where $Z^{\#}$ is the group inverse of $(I-P)$.
    \end{lemma}
    
    \subsection{Background on Submodularity and Matroids}
    \label{subsec:submod-background}
     A function $f: 2^{V} \rightarrow \mathbb{R}$ is \emph{submodular}~\cite{fujishige2005submodular} if, for any sets $S \subseteq T \subseteq V$ and any element $v \notin T$, we have $$f(S \cup \{v\}) - f(S) \geq f(T \cup \{v\}) - f(T).$$ The function $f$ is monotone if $f(S) \leq f(T)$ for $S \subseteq T$. A matroid is defined as follows. 
    \begin{definition}
    \label{def:matroid}
    Let $V$ denote a finite set and let $\mathcal{I}$ be a collection of subsets of $V$. Then $\mathcal{N} = (V,\mathcal{I})$ is a \emph{matroid} if (i) $\emptyset \in \mathcal{I}$, (ii) $S \subseteq T$ and $T \in \mathcal{I}$ implies that $S \in \mathcal{I}$, and (iii) for any $S, T \in \mathcal{I}$ with $|S| < |T|$, there exists $v \in T \setminus S$ such that $(S \cup \{v\}) \in \mathcal{I}$.
    \end{definition}
   
    The rank of a matroid $\mathcal{N}$ is equal to the cardinality of the maximal independent set in $\mathcal{I}$. A matroid basis is a maximal independent set in $\mathcal{I}$, i.e., a set $S$ such that $S \in \mathcal{I}$ and $(S \cup \{v\}) \notin \mathcal{I}$ for all $v \notin S$. A  partition matroid is defined by a partition of the set $V$ into $V=V_{1} \cup \cdots \cup V_{k}$, where $V_{i} \cap V_{j} = \emptyset$ for $i \neq j$. A set $S$ is independent in the partition matroid if, for all $i$, $|S \cap V_{i}| \leq 1$. 
    
    The following result leads to optimality bounds on local search algorithms for submodular maximization.
    
    \begin{lemma}[\cite{fisher1978analysis}]
    \label{lemma:local-opt-submodular-matroid}
    Suppose that $S$ is a basis of matroid $\mathcal{N} = (V,\mathcal{I})$, $f$ is a monotone submodular function, and there exists $\epsilon > 0$ such that, for any $u \in S$ and $v \notin S$ with $(S \cup \{v\} \setminus \{u\}) \in \mathcal{I}$, $$f(S) \geq \frac{1}{1+\epsilon}f(S \cup \{v\} \setminus \{u\}).$$ Then we have $f(S) \geq \frac{1}{2+\epsilon k}f(T)$ for any $T \in \mathcal{I}$, where $k$ is the rank of $\mathcal{N}$.
    \end{lemma}
\section{Problem Formulation and Proposed Algorithm}
\label{sec:formulation}
In this section, we first present our problem formulation, followed by the proposed algorithm.

\subsection{System Model and Problem Formulation}
\label{subsec:formulation}
We consider a  Markov Decision Process (MDP) \footnote{In this paper, we use MDP and MMDP interchangeably.} defined by a tuple $\mathcal{M} = (S, A, P, R)$, where $S$ and $A$ denote the state and action spaces, respectively. The transition probability function $P(s,a,s^{\prime})$ denotes the probability of transitioning from state $s \in S$ to state $s^{\prime} \in S$ after taking action $a \in A$. The reward function $R(s,a)$ defines the reward from taking action $a$ in state $s$.  The goal is to maximize the average reward per stage, denoted by $\lim_{T \rightarrow \infty}{\frac{1}{T}\sum_{t=0}^{T-1}{R(s_{t},a_{t})}}$, where $s_{t}$ and $a_{t}$ denote the state and action at time $t$. %We let $s_{t}^{i}$ denote the state of agent $i$ at time $t$, and let $s_{t}^{-i}$ denote the states of all agents except agent $i$ at time $t$.

The state and action spaces of $\mathcal{M}$ can be decomposed between a set of $m$ agents and an underlying environment. We write $S_{0}$ to denote the state space of the environment, $S_{i}$ the state space of agent $i$, and $A_{i}$ to denote the action space of agent $i$. We then have $S = S_{0} \times S_{1} \cdots \times S_{m}$ and $A = A_{1} \times \cdots \times A_{m}$. Throughout the paper, we use $s_{i}$ to denote a state in $S_{i}$ and $s_{-i}$ to denote a tuple of state values for the agents excluding agent $i$. Similarly, we denote an action in $A_{i}$ as $a_{i}$ and let $a_{-i}$ denote a tuple of actions for the agents excluding agent $i$. %We assume throughout that the induced Markov chain is positive recurrent

We assume that the reward function is a \emph{monotone} and \emph{submodular} function of the agent actions for any fixed state value. Define $R_{max} = \max{\{R(s,a): s \in S, a \in A\}}$ and  $R_{min} = \min{\{R(s,a): s \in S, a \in A\}}$. %A set function $f: 2^{V} \rightarrow \mathbb{R}$ is monotone if $S \subseteq T$ implies that $f(S) \leq f(T)$. A set function is \emph{submodular} if $S \subseteq T$ and $v \notin T$ implies that $f(S \cup \{v\}) - f(S) \geq f(T \cup \{v\}) - f(T)$. Intuitively, submodularity is a diminishing returns property, in which the marginal benefit from adding element $v$ to a set $S$ diminishes as the set $S$ grows. In the MDP formulation, the finite set $V = \bigcup_{i=1}^{m}{A_{i}}$.
We observe that the size of the state space may grow exponentially in the number of agents, increasing the complexity of computing the transition probabilities and optimal policy. A problem structure that is known to simplify these computations is \emph{transition independence}, defined as follows.

\begin{definition}[\cite{becker2004solving}]
\label{def:TI}
An MDP is transition independent (TI) if there exist transition functions $P_{0}:S_{0} \rightarrow S_{0}$ and $P_{i}:S_{i} \times A_{i} \rightarrow S_{i}$, $i=1,\ldots,m$, such that 
\begin{multline*}
P(\{s_{0},\ldots,s_{m}\}, \{a_{1},\ldots,a_{m}\},\{s_{0}^{\prime},\ldots,s_{m}^{\prime}\}) \\= P_{0}(s_{0},s_{0}^{\prime})\prod_{i=1}^{m}{P_{i}(s_{i},a_{i},s_{i}^{\prime})}.
\end{multline*}
\end{definition}

Transition independence implies that the state transitions of each agent depend only on that agent's states and actions, thus enabling factorization of the MDP and reducing the complexity of simulating and solving the MDP. We observe, however, that the TI property does not hold for general MDPs, and introduce the following relaxation.

\begin{definition}
\label{def:approx-TI}
Let 
\begin{multline*}
\mu_{i}(s_{i},a_{i},s_{-i},s_{-i}^{\prime},a_{-i}) \\
= Pr(s_{i}^{t+1} = \cdot | s_{t} = \{s_{i},s_{-i}\}, a_{t} = \{a_{i},a_{-i}\}, s_{-i}^{t+1} = s_{-i}^{\prime}).
\end{multline*}
An MDP is $\delta$-transition dependent (or $\delta$-dependent) if 
\begin{multline}
    \label{eq:epsilon-TI}
    \max_{\stackrel{i,s_{i},a_{i}}{\stackrel{s_{-i},s_{-i}^{\prime}, s_{-i}^{\prime\prime}}{s_{-i}^{\prime\prime\prime},a_{-i},a_{-i}^{\prime}}}}{||\mu_{i}(s_{i},a_{i},s_{-i},s_{-i}^{\prime},a_{-i})} \\
    -\mu_{i}(s_{i},a_{i},s_{-i}^{\prime\prime},s_{-i}^{\prime\prime\prime},a_{-i}^{\prime})||_{TV} \leq \delta
\end{multline}

   % ||P(s_{t+1}^{i} = \cdot | s_{t} = \{s_{i},s_{-i}\}, a_{t} = \{a_{i},a_{-i}\}, s_{t+1}^{-i}=s_{-i}^{\prime})} \\
  %  -P(s_{t+1}^{i} = \cdot | s_{t} = \{s_{i},s_{-i}^{\prime\prime}\}, a_{t}=\{a_{i},a_{-i}^{\prime}\},s_{t+1}^{-i}=s_{-i}^{\prime\prime\prime})||_{TV} \leq \epsilon
   % \max_{i,s_{i},a_{i},a_{-i},a_{-i}^{\prime},s_{-i},s_{-i}^{\prime},s_{-i}^{\prime\prime},s_{-i}^{\prime\prime\prime}}{||P(s_{t+1,i}|s_{t}=s_{i},a_{t,i}=a_{i},a_{t,-i}=a_{-i},s_{t,-i}=s_{-i},s_{t+1,-i}=s_{-i}^{\prime})-P(s_{t+1,i} | s_{t}=s_{i},a_{t,i}=a_{i},a_{t,-i}=a_{-i}^{\prime},s_{t,-i} = s_{-i}^{\prime\prime},s_{t+1,-i}=s_{-i}^{\prime\prime\prime})||_{TV}} \leq \epsilon
%\end{multline}
\end{definition}
Intuitively, the $\delta$-dependent property implies that the impact of the other agents on agent $i$'s transition probabilities is bounded by $\delta$. When $\delta=0$, our definition of $\delta$-dependent MDP reduces to TI-MDP defined in \cite{becker2004solving}.

The agents choose their actions at each time step by following a policy, which maps the current and previous state values to the action at time $t$. We focus on \emph{stationary policies} of the form $\pi: S \rightarrow A$, which only incorporate the current state value when choosing the next action. We assume that, for any stationary policy, the resulting induced Markov chain is ergodic. We let $\overline{\lambda}$ denote the maximum value of the ergodic number $\Lambda_{1}(Z^{\#})$ over all stationary policies. Furthermore, to reduce the complexity of storing the policy at the agents, each agent follows a \emph{local} policy $\pi_{i}: S_{0} \times S_{i} \rightarrow A_{i}$. Hence, each agent's actions only depend on the environment and the agent's internal state. Any policy $\pi$ with this structure can be expressed as $\{\pi_{1},\ldots,\pi_{m}\}$, where $\pi_{i}$ denotes the policy of agent $i$. We let $\pi_{-i}$ denote the set of policies of the agents excluding $i$.

The problem is formulated as follows. Define the value function for policies $\{\pi_{i} : i=1,\ldots,m\}$ by $$J(\pi) = \lim_{T \rightarrow \infty}{\mathbf{E}\left\{\frac{1}{T}\sum_{t=0}^{T-1}{R(s_{t},\pi(s_{t}))}\right\}}.$$  When it is not clear from the context, we let $J_{\mathcal{M}}(\pi)$ denote the average reward from policy $\pi$ on MDP $\mathcal{M}$. The goal is then to select $\pi$ that maximizes $J(\pi)$. As a preliminary, we say that a policy $\pi$ is locally optimal if, for all $i$, $J(\pi) \geq J(\tilde{\pi}_{i}, \pi_{-i})$ for any agent $i$ policy $\tilde{\pi}_{-i}$. We say that $\pi$ is $\theta$-locally optimal if $(1+\theta)J(\pi) \geq J(\tilde{\pi}_{i}, \pi_{-i})$ for all $i$ and all policies $\tilde{\pi}_{i}$ for agent $i$.

\subsection{Proposed Algorithm}
\label{subsec:algo}

To motivate our approach, we first map the problem to a combinatorial optimization problem as in \cite{kumar2017decentralized}. Consider the finite set of \emph{agent policies}, which we write as $\Pi = \Pi_{1} \cup \cdots \cup \Pi_{m}$, where $\Pi_{i}$ denotes the set of possible local policies for agent $i$. The collection $\Pi_{i}$ is formally defined as the set of functions of the form $\{\pi_{i}: S_{0} \times S_{i} \rightarrow A_{i}\}$. The problem of selecting an optimal collection of local policies can therefore be mapped to the combinatorial optimization problem 
\begin{equation}
    \label{eq:combinatorial-form}
    \begin{array}{ll}
\mbox{maximize} & J(\pi) \\
\mbox{s.t.} & \pi \in \Pi, |\pi \cap \Pi_{i}| = 1 \ \forall i=1,\ldots,m
\end{array}
\end{equation}
In (\ref{eq:combinatorial-form}), the policy $\pi$ is interpreted as a set, in which each element represents the policy of a single agent. Since there is exactly one policy per agent, the constraint $|\pi \cap \Pi_{i}| =1$ is a partition matroid constraint. The following proposition provides additional structure for a special case of (\ref{eq:combinatorial-form}).
\begin{proposition}[\cite{kumar2017decentralized}]
\label{prop:submodular-TI}
If the MDP $\mathcal{M}$ is transition-independent and the rewards $R(s,a)$ are monotone and submodular in $a$ for any fixed state $s$, then the function $J_{\mathcal{M}}: \Pi \rightarrow \mathbb{R}$ is monotone and submodular in $\pi$.
\end{proposition}

Proposition \ref{prop:submodular-TI} implies that, when the MDP is TI and reward function is submodular, efficient heuristic algorithms will lead to provable optimality guarantees. One such algorithm is local search, which attempts to improve the current set of policies $\{\pi_{1},\ldots,\pi_{m}\}$ by searching for policies $\pi_{i}^{\prime}$ satisfying $J(\pi_{i}^{\prime}, \pi_{-i}) > J(\pi)$. If no such policy can be found, then the policy $\pi$ is a local optimum of (\ref{eq:combinatorial-form}), and hence Lemma \ref{lemma:local-opt-submodular-matroid} can be used to obtain a $\frac{1}{2}$-optimality guarantee.

The difficulty in the above approach arises from the fact that the number of possible policies $\Pi_{i}$ for each agent $i$ grows exponentially in the number of states $S_{i}$. Hence, instead of brute force search, the approach of \cite{kumar2017decentralized} leverages the fact that, in a TI-MDP in which all other agents adopt stationary policies,  the optimal policy for agent $i$ can be obtained as the solution to an MDP.  This MDP has reward function and transition matrix, respectively, given by 
\begin{equation*}
R_{i}(s_{i},a_{i}) = \sum_{s_{-i}}{q(s_{-i})R(\{s_{i},s_{-i}\}, \{a_{i},\pi_{-i}(s_{-i})\})}
\end{equation*}
and $P_{i}(s_{i},a_{i},s_{i}^{\prime}) = P_{i}$, where $q$ denotes the stationary distribution of the joint states under the chosen policies. 
Using this property, an optimal policy for agent $i$, conditioned on the policies $\{\pi_{-i}\}$ of the other agents, can be obtained by solving this equivalent MDP. 

%{\color{red}
We now present our proposed approach, which generalizes this idea from TI to non-TI MDPs.  Our algorithm is initialized as follows. Choose a parameter $\epsilon > 0$. First, for each agent $i$, choose a probability distribution $\mu_{i}$ over the states in $S_{-i}$ and a policy $\pi_{-i}(s_{-i})$. Next,  define a local transition function for each agent $i$ as 
\begin{multline}
    \label{eq:algo-local-transition-function}
P_{i}(s_{i},a_{i},s_{i}^{\prime}) \\
= \mathbf{E}_{\mu_{i}}(P(\{s_{i},s_{-i}\}, \{a_{i},\pi_{i}(s_{-i})\}, \{s_{-i}^{\prime}, s_{-i}^{\prime}\})),
\end{multline}
where the expectation is over $s_{-i}$ from distribution $\mu_{-i}$. 
We then choose policies $\pi_{1}^{0},\ldots,\pi_{m}^{0}$ arbitrarily, and set $\hat{q}_{i}^{0}$ as the stationary distribution on the state $s_{i}$ induced by the policy $\pi_{i}^{0}$ under transition function $P_{i}^{0}$.

At the $k$-th iteration of the algorithm, each agent $i$ updates its policy $\pi_{i}$ while the other agent policies are held constant. The optimal policy of agent $i$ is approximated by the solution to a local MDP denoted $\mathcal{M}_{i}^{k} = (S_{i},A_{i},P_{i},R_{i}^{k})$, where 
\begin{multline}
\label{eq:algo-local-reward}
R_{i}^{k}(s_{i},a_{i})\\
= \sum_{s_{-i}}{\left[\left(\prod_{j \neq i}{\hat{q}_{j}(s_{j})}\right)R(\{s_{i},s_{-i}\},\{a_{i},\pi_{-i}(s_{-i})\})\right]}.
\end{multline}
A policy $\pi_{i}$ is then obtained as the optimal policy for $\mathcal{M}_{i}^{k}$. If $J_{\mathcal{M}_{i}^{k}}(\pi_{i}) \geq (1+\epsilon)J_{\mathcal{M}_{i}^{k}}(\pi_{i}^{k-1})$, then set $\pi_{i}^{k}$ equal to $\pi_{i}^{k-1}$, compute $\hat{q}_{i}^{k}$ as the stationary distribution of $P_{i}$ under policy $\pi_{i}^{k}$, and increment $k$. The algorithm terminates when no agent modifies its policy in an iteration $k$.

Pseudocode for this algorithm is given in Algorithm \ref{algo:Stack}. 

%{\color{red}
%\textbf{Remarks on this new algorithm:} This algorithm should improve on the complexity of the algorithm from the AAAI paper for the following reasons:
%\begin{itemize}
 %   \item There is no longer a need to compute the joint stationary probability distribution of the agents, which was the source of the exponential complexity in the AAAI version. In other words, the Monte Carlo step from the AAAI paper can be \emph{completely eliminated}.
  %  \item There is a potentially exponential complexity in the reward function computation in (\ref{eq:algo-local-reward}). However, this reward function is probably easier to approximate via Monte Carlo. Moreover, I believe that for certain reward functions, including the ones in our simulation section, there is a \emph{closed form} for the reward function in (\ref{eq:algo-local-reward}).
   % \item I think that the optimality bounds for this algorithm should be \emph{similar} to the algorithm in the AAAI version.
    %\item The only potential drawback is that the empirical performance (reward) may not be as high, however, we will need to do some simulations to check this.
%\end{itemize}

%}

\begin{center}
	\begin{algorithm}[h]
		\caption{Approximate algorithm for selecting local policies}
		\label{algo:Stack}
		\begin{algorithmic}[1]
			\State \textbf{Input:} MDP $(S,A,P,R)$
			\State \textbf{Output:} Policies $\{\pi_{i}: i=1,\ldots,m\}$
			\State \textbf{Initialization:} $\pi_{i}^{0}(s_{i}) \leftarrow \frac{1}{|A_{i}|}$, $i = 1,\ldots,m$, $s_{i} \in S_{i}$, $k \rightarrow 1$
			\State Compute $P_{i}: i=1,\ldots,m$ according to (\ref{eq:algo-local-transition-function})
			\State Compute $\hat{q}_{i}:i=1,\ldots,m$ as the stationary distribution of $P_{i}$ under $\pi_{i}^{0}$
			%\State \textbf{Initialization: } $n \leftarrow 0$, $v^{0}_{k} \leftarrow 0$, $\rho^{0}_{k} \leftarrow 0$, $\epsilon^{0}_{k} %\leftarrow 0$, $\pi^{0}_{k}~\leftarrow~\boldsymbol{\pi}_{k}$ for $k \in \{D, A\}$ and $s \leftarrow s_0$.
			\While {1}
			\State $\pi_{i}^{k} \leftarrow \pi_{i}^{k-1}$, $i=1,\ldots,m$, $found \leftarrow 0$, $\hat{q}_{i}^{k} \leftarrow \hat{q}_{i}^{k-1}$, $i=1,\ldots,m$
			\For{$i=1,\ldots,m$}
			\State Compute $R_{i}^{k}$  according to  (\ref{eq:algo-local-reward}).
			\State Solve local MDP $\mathcal{M}_{i}^{k}=(S_{i},A_{i},P_{i},R_{i}^{k})$ to obtain new policy $\pi_{i}$.
			\If{$J_{\mathcal{M}_{i}^{k}}(\pi_{i}) > (1+\epsilon)J_{\mathcal{M}_{i}^{k}}(\pi_{i}^{k-1})$}
			\State $\pi_{i}^{k} \leftarrow \pi_{i}$
			\State $\hat{q}_{i}^{k} \leftarrow$ stationary distribution of $P_{i}$ under policy $\pi_{i}$
			\State $found \leftarrow 1$; \textbf{Break}
			\EndIf
			\EndFor 
			\If{$found == 0$}
			\State \textbf{Break}
			\Else
			\State $k \leftarrow k+1$
			\EndIf
			\EndWhile
		\end{algorithmic}\label{algo}
	\end{algorithm}
	\vspace*{-3 mm}
\end{center}

\section{Optimality Analysis}
\label{sec:optimality}
%This section considers the optimality of our proposed approach. %We first analyze the optimality guarantees for an $\epsilon$-TI MDP. We then characterize the parameter $\epsilon$. Then, we analyze the complexity of our approach.

%\subsection{Optimality of Algorithm}
%\label{subsec:optimality}
We analyze the optimality in three stages. First, we define a TI-MDP, and prove that the policies returned by our algorithm are within a provable bound of a local optimum of the TI-MDP. We then use submodularity of the reward function to prove that the local optimal policies provide a constant-factor approximation to the global optimum on the TI-MDP. Finally, we prove that the approximate global optimum on the TI-MDP is also an approximate global optimum for the original MDP.

%{\color{red}
We define $\hat{\pi} = \{\hat{\pi}_{1},\ldots,\hat{\pi}_{m}\}$ to be the policy returned by our algorithm. Let $\hat{q}$ be the joint stationary distribution of the agents in the MDP $\mathcal{M}$ arising from these policies. We construct a TI-MDP $\hat{\mathcal{M}} = (S, A, \hat{P}, \hat{R})$. The transition function $\hat{P}$ is defined by $$\hat{P}(s,a,s^{\prime}) = \prod_{i=1}^{m}{\hat{P}_{i}(s_{i},a_{i},s_{i}^{\prime})}.$$ The reward function $\hat{R} = R(s,a)$. We observe that, by construction, if $\mathcal{M}_{i}$ is the local MDP obtained at the last iteration of Algorithm \ref{algo:Stack}, then $J_{\mathcal{M}_{i}}(\hat{\pi}_{i}) = J_{\hat{\mathcal{M}}}(\hat{\pi})$ for all $i$. 

\begin{lemma}
\label{lemma:TI-MDP-local-opt}
The policy $\hat{\pi}$ returned by Algorithm \ref{algo:Stack} is a $1/(1+\epsilon)$-local optimum for MDP $\hat{\mathcal{M}}$.
\end{lemma}

\emph{Proof:} By construction, the algorithm terminates if, for all $i$, there is no policy $\pi_{i}$ such that $$J_{\hat{\mathcal{M}}}(\pi_{i}, \hat{\pi}_{-i}) = J_{\mathcal{M}_{i}}(\pi_{i}) \geq (1+\epsilon)J_{\mathcal{M}_{i}}(\hat{\pi}_{i}) = (1+\epsilon)J_{\hat{\mathcal{M}}}(\hat{\pi}),$$ implying that $\hat{\pi}$ is a $1/(1+\epsilon)$-local optimum of $\hat{\mathcal{M}}$.

Based on the local optimality, we can derive the following optimality bound for $\hat{\pi}$.
\begin{lemma}
\label{lemma:TI-MDP-optimality}
Let $\pi^{\ast}$ denote the optimal local policies for MDP $\mathcal{M}$. Then $$J_{\hat{\mathcal{M}}}(\hat{\pi}) \geq \frac{1}{2+\epsilon  m}J_{\hat{\mathcal{M}}}(\pi^{\ast}).$$
\end{lemma}

\emph{Proof:} The proof follows from the submodularity of $J_{\hat{\mathcal{M}}}$ (Proposition \ref{prop:submodular-TI}) and Lemma \ref{lemma:local-opt-submodular-matroid}.

Lemma \ref{lemma:TI-MDP-optimality} provides an optimality bound with respect to the TI-MDP $\hat{\mathcal{M}}$. Next, we leverage the $\delta$-dependent property to derive an optimality bound with respect to the given MDP $\mathcal{M}$. We start with the following  preliminary results.

\begin{lemma}
\label{lemma:total-variation-bound}
For any state $s$ and policy $\pi$, $||P(s,\pi,\cdot)-\overline{P}(s,\pi,\cdot)||_{TV} \leq m\delta$, where $P$ and $\overline{P}$ are the transition matrices corresponding to $\mathcal{M}$ and $\hat{\mathcal{M}}$, respectively.
\end{lemma}

A proof can be found in the technical appendix. We next exploit the bound in Lemma \ref{lemma:total-variation-bound} to approximate the gap between the $J_{\mathcal{M}}$ and $J_{\hat{\mathcal{M}}}$.

\begin{lemma}
\label{lemma:MDP-approximation}
Suppose that $\mathcal{M}$ is $\delta$-dependent MDP. For any policy $\pi$, $|J_{\hat{\mathcal{M}}}(\pi) - J_{\mathcal{M}}(\pi)| \leq (R_{max}-R_{min})(2\overline{\lambda}m\delta)$. 
%Define $\phi = \frac{R_{max}}{R_{min}}2m\epsilon\overline{\lambda}$.  For any policy $\pi$, $J_{\hat{\mathcal{M}}}(\pi) \geq (1-\phi)J_{\mathcal{M}}(\pi)$ and $J_{\mathcal{M}}(\pi) \geq (1-\phi)J_{\hat{\mathcal{M}}}(\pi)$.
\end{lemma}

\emph{Proof:} Let $q_{\mathcal{M}}$ and $q_{\hat{\mathcal{M}}}$ denote the stationary distributions induced by policy $\pi$ on MDPs $\mathcal{M}$ and $\hat{\mathcal{M}}$. We have $$|J_{\hat{\mathcal{M}}}(\pi)-J_{\mathcal{M}}(\pi)| \leq (R_{max}-R_{min}) ||q_{\mathcal{M}}-q_{\hat{\mathcal{M}}}||_{1}.$$ By Lemma \ref{lemma:perturbation}, $$||q_{\mathcal{M}}-q_{\hat{\mathcal{M}}}||_{1} \leq 2\overline{\lambda}m\delta,$$  giving the desired result.

Combining these derivations yields the following.

\begin{theorem}
\label{theorem:optimality-bound}
Let $\mathcal{M}$ be an $\delta$-dependent MDP and $\hat{\pi}$ and $\pi^{\ast}$ denote the output of Algorithm \ref{algo:Stack} and the optimal policies, respectively. Then $$J_{\mathcal{M}}(\pi^{\ast}) \leq 4R_{max}\overline{\lambda}m\delta + (1+m\epsilon)J_{\hat{\mathcal{M}}}(\hat{\pi}) + J_{\mathcal{M}}(\hat{\pi}).$$
\end{theorem}

\emph{Proof:} We have 
\begin{IEEEeqnarray*}{rCl}
\IEEEeqnarraymulticol{3}{l}{
J_{\mathcal{M}}(\pi^{\ast}) - J_{\mathcal{M}}(\hat{\pi})}\\
&\leq& |J_{\mathcal{M}}(\hat{\pi})-J_{\hat{\mathcal{M}}}(\hat{\pi}) + J_{\hat{\mathcal{M}}}(\hat{\pi}) - J_{\hat{\mathcal{M}}}(\pi^{\ast})| \\
&& + |J_{\hat{\mathcal{M}}}(\pi^{\ast}) - J_{\mathcal{M}}(\pi^{\ast})| \\
&\leq& 4R_{max}\overline{\lambda}m\delta + (1+m\epsilon)J_{\hat{\mathcal{M}}}(\hat{\pi})
\end{IEEEeqnarray*}
by Lemmas \ref{lemma:TI-MDP-optimality} and \ref{lemma:MDP-approximation}.

\section{Simulation}\label{sec:simulation}

\begin{table*}[t!]
\centering
\begin{adjustbox}{width=1\textwidth}
\begin{tabular}{|c|c|c|c|c|c|c|c|c|c|c|c|}
  \hline
     \multicolumn{6}{|c|}{MDP Setting} & \multicolumn{3}{|c|}{Average Reward } & \multicolumn{3}{|c|}{Run Time (seconds)} \\
    \hline
    \thead{Number of Agents\\ and Targets $(N,B)$} & \thead{Grid Size\\ $L$}  & Target Location & \thead{Agents' Initial\\ Locations}&  \thead{State Space\\ Size} & \thead{Action Space\\ Size} &Algorithm \ref{algo:Stack} & \thead{Global MDP\\ Approach} & Ratio & Algorithm \ref{algo:Stack} & \thead{Global MDP\\ Approach} & Ratio \\
    \hline
  2, 1& $3 \times 3$ & 6 & (0,2) & 81 & 16 & 0.312 & 0.333 & 93.69\% & 0.246& 0.142 & 173.2\% \\
  \hline
  2, 2 & $5 \times 5$ & (20,24) & (3,5) & 625 & 16 & 0.268 & 0.269 & 99.63\% & 3.08 & 3.01 & 102.3\% \\
    \hline
  3, 1 & $3 \times 3$ & 6 & (0,0,2) & 729 & 64  & 0.460 & 0.504 & 91.27\% & 6.51  & 24.1 & 27.01\%\\
    \hline
    3, 1 & $3 \times 3$ & 8 & (1,1,2) & 729 & 64  & 0.435 & 0.475 & 91.58\% & 6.66  & 25.2 & 26.43\%\\
    \hline
    3, 1 & $4 \times 4$ & 15 & (0,0,3) & 4096 & 64 & 0.318 & 0.334 & 95.21\% & 41.9 & 446 & 9.395\%\\
    \hline
     3, 1 & $4 \times 4$ & 12 & (1,1,2) & 4096 & 64 & 0.337 & 0.355 & 94.93\% & 41.9 & 439 & 9.544\%\\ 
    \hline
    4, 1 & $2 \times 2$ & 3 & (0,0,1,1) & 256 & 256 & 0.758 & 0.766  & 98.96\% & 4.76 & 35.7 & 13.33\% \\
    \hline
    2, 2 & $10 \times 10$ & (90,99) & (0,9) & 10000 & 16 & 0.125 & 0.125 & 100\% & 56.9 &  451 & 12.62\% \\
    \hline
    2, 2 & $10 \times 10$ & (55,77) & (5,99) & 10000 & 16 & 0.241 & 0.241 & 100\% & 56.8 &  429 & 13.24\% \\
    \hline
\end{tabular}
\end{adjustbox}
\caption{Comparison of average reward and run time when Multi-robot control problem is solved using Algorithm \ref{algo:Stack} versus using relative value iteration algorithm in \textit{Python MDP Toolbox} \cite{MDPtoolbox} on corresponding global MDP ($\mathcal{M}$). Both appraches are compared under different sizes of grids $L$, number of agents $N$, number of targets $B$, and initial locations of the agents. The `Ratio' of average reward in Table \ref{table:robot} is obtained as the average reward of Algorithm \ref{algo:Stack} divided by that of the global MDP approach. The `Ratio' of run time in Table \ref{table:robot} is obtained as the run time incurred by Algorithm \ref{algo:Stack} divided by that of the global MDP approach.} 
\label{table:robot}
\end{table*}

\begin{table*}[t!]
\centering
\begin{adjustbox}{width=1\textwidth}
\begin{tabular}{|c|c|c|c|c|c|c|c|c|c|}
  \hline
     \multicolumn{4}{|c|}{MDP Setting} & \multicolumn{3}{|c|}{Average Reward} & \multicolumn{3}{|c|}{Run Time (seconds)}  \\
    \hline
    \thead{Patrol Unit, Adversary\\Number} & \thead{Location\\ Number} & \thead{State Space\\ Size} & \thead{Patrol Units' Action \\Space Size} &Algorithm \ref{algo:Stack} & \thead{Global MDP\\Approach} &Ratio &Algorithm \ref{algo:Stack} & \thead{Global MDP\\Approach} &Ratio \\
    \hline
   2,1& 3 & 27 & 9 & 0.774 & 0.775 & 99.87\% & 0.133 & 0.435 & 30.57\% \\
   \hline
   3,1 & 3 & 81 & 27 & 0.865 & 0.866 & 99.88\% & 1.93 & 6.49 & 29.73\% \\
    \hline
   3,2 & 3 & 243 & 27  & 1.73 & 1.73 & 100\% & 22.84  & 157 & 14.54\% \\
    \hline
    2,1 & 5 & 125 & 25 & 0.768 & 0.768 & 100\% & 5.56 & 18.13 & 30.67\% \\
    \hline
    3,1 & 5 & 625& 125 & 0.856 & 0.856 & 100\% & 239 &  2823 & 8.466\% \\
    \hline
    2,1 & 7 & 343& 49 & 0.766 & 0.766 & 100\% & 68.7 &  358 & 19.19\% \\
    \hline
    2,1 & 8 & 512& 64 & 0.766 & 0.766 & 100\% & 188 & 1295 & 14.52\% \\
    \hline
\end{tabular}
\end{adjustbox}
\caption{Comparison of average reward and run time when Multi-Agent Patrolling example is solved using Algorithm \ref{algo:Stack} versus using relative value iteration algorithm in \textit{Python MDP Toolbox} \cite{MDPtoolbox} on corresponding global MDP ($\mathcal{M}$). Both approaches are compared under different number of patrol units, adversaries, and locations. The `Ratio' of average reward in Table \ref{table:robot} is obtained as the average reward of Algorithm \ref{algo:Stack} divided by that of the global MDP approach. The `Ratio' of run time in Table \ref{table:robot} is obtained as the run time incurred by Algorithm \ref{algo:Stack} divided by that of the global MDP approach.} 
\label{table:security}
\end{table*}

In this section, we present our simulation results. We consider two scenarios, namely, multi-robot control and a multi-agent patrolling example. Both simulations are implemented using Python 3.8.5 on a workstation with Intel(R) Xeon(R) W-2145 CPU @ 3.70GHz processor and $128$~GB memory. Given the transition and reward matrices, an MDP is solved using \textit{Python MDP Toolbox} \cite{MDPtoolbox}. 

\subsection{Multi-robot Control}
\subsubsection{Simulation Settings}
We consider a set of $N > 1$ robots whose goal is to cover maximum number of targets from a set of fixed targets $B$ positioned in a $L \times L$ grid environment. Robots initially start from a fixed set of grid locations. At each time $t = 1, 2, \ldots$, each robot $i$ can move one grid position horizontally or vertically from the current grid position by taking some action $a_i\in A_i := \{\mbox{left, down, right, up}\}$. For each robot $i$, let $D_i := \{d(a_i)\}_{a_i \in A_i}$ denotes the set of grid positions that can be reached from the current grid position $s_i$ under each action $a_i \in A_i$. $d(a_i)$ is the grid position corresponds to $(s_i, a_i)$. Note that $d(a_i) = \varnothing$, if $a_i$ is not a valid action (e.g., $a_i \in \{left, down\}$ at the bottom left corner of the grid are not valid actions). Let $P_i$ be the transition probability function associated with robot $i$ and $P_i(s_i, a_i, s_i')$ be the probability of robot $i$ transitions from a grid position $s_i$ to $s_i'$ under action $a_i$. Let $n(s_i')$ be the number of robots at $s_i'$ after taking actions $(a_i, a_{-i})$. Then,
\begin{equation*}
    P_i(s_i,a_i,s_i')=\begin{cases}
    c,~\mbox{if $s_i' = d(a_i)$ and $n(s_i') < K$}\\
    \frac{1-c}{|D_i|-1},\mbox{if $s_i' \in D_i \setminus d(a_i)$ and $n(s_i') < K$}\\
    \delta c,~\mbox{if $s_i' = d(a_i)$ and $n(s_i') \geq K$}\\
    \frac{1-\delta c}{|D_i|-1},\mbox{if $s_i' \in D_i \setminus d(a_i)$ and $n(s_i') \geq K$}\\
    0,~\mbox{otherwise, }
    \end{cases}
\end{equation*}
where $K \geq 1$. Uncertainty in the environment is modeled by the parameter $0 \leq c \leq 1$ and the  transition dependencies between the robots are modeled by the parameter $0 \leq \delta\ \leq 1$.

We model the multi-robot control problem as an MDP $\mathcal{M} = (S, A, P, R)$. The state space $S = S_1 \times \ldots \times S_N$, where $S_i=\{0, \ldots, L^2-1\}$ for all $i=1,\ldots,N$. The action space $A = A_1 \times \ldots \times A_N$. The transition probability matrix is denoted as $P$. The probability of transitioning from a state $s=(s_1,\ldots,s_N)\in S$ to some target state $s'=(s_1',\ldots,s_N')\in S$ by taking action $a=(a_1,\ldots,a_m)\in A$ is given as $P(s,a,s')=\prod_{i=1}^mP_i(s_i,a_i,s_i')$. The \textit{submodular} reward $R$ of $\mathcal{M}$ is given by $R(s_{t},\pi(s_{t}))= \sum_{b \in B}(1-(1-\eta)^{N_b})$, where $N_b$ is the number of robots visiting target $b$ following a joint policy $\pi(s_{t})$ at a state state $s_{t}$. The parameter $\eta$ captures the effectiveness of having $N_b$ agents at target $b$. Similar \textit{submodular} reward has been used in \cite{kumar2017decentralized}.

%  The parameter $\epsilon$ in this problem setting defines the transition dependency in the transition probability matrix $P$ of $\mathcal{M}$. Specifically, when $\epsilon \rightarrow 1$, the transition dependency among agents becomes weaker, and MDP $\mathcal{M}$ approaches TI-MDP. Hence, this formulation defines an $\epsilon$-TI MDP $\mathcal{M}$.

\subsubsection{Simulation Results}

We use Algorithm~\ref{algo:Stack} to find a set of policies for the robots that maximizes their average reward. Parameters $\epsilon$, $K$, $\eta$, and $\beta$ are set to $0$, $1$, $0.75$ and $0.9$, respectively. The transition probability for each agent is calculated by evaluating \eqref{eq:algo-local-transition-function} over $\lfloor\frac{N*L*L}{2}\rfloor$ samples of actions $\{a_i,\Bar{a}_{-i}\}$ and states $\{s_{i}',\Bar{s}_{-i}'\}$. We test Algorithm \ref{algo:Stack} under different sizes of grids $L$, number of agents $N$, number of targets $B$, and initial locations of the agents. For each setting, we execute Algorithm \ref{algo:Stack} for $100$ trials, and take the average over the trials as the performance of Algorithm~\ref{algo:Stack}. We compare Algorithm \ref{algo:Stack} with the a global MDP approach, which calculates the optimal values using relative value iteration algorithm \cite{bertsekas1995dynamic} provided by \textit{Python MDP Toolbox} \cite{MDPtoolbox} on MDP $\mathcal{M}$. Note that the state space of MDP $\mathcal{M}$ is exponential in grid size $L$ and number of agents $N$. The action space is exponential in $N$. The total size of all local MDPs for all agents constructed using \eqref{eq:algo-local-transition-function} and \eqref{eq:algo-local-reward} grows linearly with respect to the number of agents. As the number of agents and/or grid size of examples increase, the global MDP approach incurs a heavy memory computation overhead to the system. For an example, when $2$ robots trying to reach $1$ target in a $10 \times 10$ grid, it requires around $1.29$~Gb of memory to compute the solution using global MDP approach, while our proposed approach only requires $1.45$~Mb memory to calculate the policy. Therefore, the global MDP approach is not computationally efficient for larger example sizes.

Table \ref{table:robot} shows the simulation results obtained using Algorithm \ref{algo:Stack} and the global MDP approach. We observe that our proposed approach provides more than $90\%$ optimality with respect to the average reward achieved by the agents for all settings, while incuring comparable run time when the example size is small and much less run time when the example sizes increase. Particularly, as the number of agents and the grid size increase, e.g., two agents, two targets, and $10\times 10$ grid, our proposed approach maintains more than $100\%$ optimality with only $12.62\%$ run time, compared with the global MDP approach. Hence, our proposed approach shows scalability to mutli-agent scenarios with $\delta$-dependent property.

\subsection{Multi-Agent Patrolling Example}
\subsubsection{Simulation Settings}
We implement our proposed approach on a patrolling example with multiple patrol units capturing multiple adversaries among a finite set of locations $L$ as an evaluation. At each time, each patrol unit can be deployed at some location $l\in L$.

The objective of the patrol units is to compute a policy to patrol the locations to capture the adversaries. Each adversary is assumed to follow a heuristic policy as follows. If there exists no patrol unit that is deployed at the adversary's target location $l$, then with probability $d$ the adversary transitions to location $l$ and with probability $(1-d)/(L-1)$ the adversary transitions to some other location $l'\neq l$. If the adversary's target location $l$ is being patrolled by some unit, then with probability $\beta d$ the adversary transitions to location $l$, and with probability $(1-\beta d)/(L-1)$ the adversary transitions to some other location $l'\neq l$. The adversaries' policies are assumed to be known to the patrol units.

The patrolling example is modeled by an MDP $\mathcal{M}=(S,A,P,R)$, where $S=(\times_iS_i)\times(\times_jS_j)$ is the set of joint locations of the patrol units and adversaries, with $S_i=L$ is the set of locations at which patrol unit $i$ is deployed and $S_j=L$ is the set of locations where adversary $j$ can be located. The action set of each patrol unit and adversary is $A_i=A_j=L$. Thus the joint action space $A=(\times_iA_i)\times(\times_jA_j)$. We shall remark that the joint action space is defined as the Cartesian product of the action spaces of all patrol units and adversaries so that we can accurately capture the transition probabilities of all patrol units and adversaries. We solve the problem by optimizing over the joint action space of all the patrol units, since the adversaries' policies are known to the patrol units. For each patrol unit $i$ and adversary $j$, we let
\begin{align*}
    &P_i(s_i, a_i,s_i') =\begin{cases}
    c&\mbox{ if } a_i=s_i', \not\exists i'\neq i \text{ s.t. }a_{i}=a_{i'}\\
    \frac{1-c}{|L|}&\mbox{ if }a_i\neq s_i',\not\exists i'\neq i \text{ s.t. }a_{i}=a_{i'}\\
    \delta c&\mbox{ if } a_i=s_i', \exists i'\neq i \text{ s.t. }a_{i}=a_{i'}\\
    \frac{1-\delta c}{|L|}&\mbox{ if } a_i\neq s_i', \exists i'\neq i \text{ s.t. }a_{i}=a_{i'}
    \end{cases}\\
    &P_j(s_j, a_j,s_j') =\begin{cases}
    d&\mbox{ if } a_j=s_j', \not\exists i \text{ s.t. }a_{i}=a_{j'}\\
    \frac{1-d}{|L|}&\mbox{ if }a_j\neq s_j', \not\exists i \text{ s.t. }a_{i}=a_{j'}\\
    \beta d&\mbox{ if } a_j=s_j', \exists i \text{ s.t. }a_{i}=a_{j'}\\
    \frac{1-\beta d}{|L|}&\mbox{ if } a_j\neq s_j',\exists i \text{ s.t. }a_{i}=a_{j'}
    \end{cases}
\end{align*}
Here parameters $c,d\in[0,1]$ capture the transition uncertainties, parameter $\delta\in[0,1]$ captures the transition dependency among the patrol units, and $\beta\in[0,1]$ captures the adversaries' reactions to the patrol units' actions. Let $s=(\times_is_i)\times(\times_js_j)$ and $s'=(\times_is'_i)\times(\times_js'_j)$ be two joint locations. Then $P(s,a,s')=\prod_{i,j}P_i(s_i,a_i,s_i')P_j(s_j,a_j,s_j')$, where $a=(\times_is_i)\times(\times_js_j)$. We define the reward function $R(s,a)$ for each $s\in S$ and $a\in A$ as $R(s,a)=\sum_{s'\in S}r(s,a,s')P(s,a,s')$, where $r(s,a,s')=\sum_{l\in L}(1-(1-\eta)^{k_l})x_l$, where $\eta\in(0,1]$ is the effectiveness parameter, $k$ and $x$ are the number of patrol units and adversaries that are in location $l$ corresponding to $s'$, respectively.

% Given the current  location $s$, each defender $i$ computes its policy $\pi_i:S\mapsto L$ to determine the next location where its resource will be deployed. Each adversary $j$ determines its next location by following policy $\tau_j:S\mapsto L$. We assume that the policy of each adversary $j$ is known to the defenders. 

\subsubsection{Simulation Results}

We use Algorithm \ref{algo:Stack} to compute the policies for the patrol units, given the adversaries' policies. Parameters $c$, $d$, $\delta$, $\beta$, and $\eta$ are set as $0.9$, $1$, $0.9$, $0.9$, and $0.75$, respectively. We calculate the transition probability of each patrol unit $i$ by evaluating \eqref{eq:algo-local-transition-function} over all possible actions $(\times_{-i}a_{-i})\times(\times_{j}a_{j})$ and all possible states $(\times_{-i}s_{-i})\times(\times_{j}s_{j})$ of all adversaries and all the other patrol units except $i$. We implement our proposed approach under various settings by varying the number of patrol units, adversaries, and locations. For each setting, we run Algorithm \ref{algo:Stack} for $100$ trials and take the average over the trials as its performance. We compare Algorithm \ref{algo:Stack} with the global MDP approach that implements relative value iteration algorithm on MDP $\mathcal{M}$.

Table \ref{table:security} shows the simulation results obtained using Algorithm \ref{algo:Stack} and the global MDP approach. We observe that our proposed approach achieves more than $99\%$ of optimality with respect to the average reward, while incuring at most $30.57\%$ of run time over all settings. By comparing the first row, 4-th row, 6-th row, and 7-th row in Table \ref{table:security}, we have that the run time advantage provided by our proposed approach remains when we increase the number of locations. By comparing the first three rows in Table \ref{table:security}, we observe that our proposed approach remains close to optimal average reward (more than $99\%$), but scales better when the number of agents including patrol units and adversaries increases.

\section{Conclusions}
\label{sec:conclusion}
This paper presented an approach for selecting decentralized policies for transition dependent MMDPs. We proposed a property of $\delta$-transition dependence, which we defined based on the maximum total variation distances for each agent's state transitions conditioned on the actions of the other agents. In the special case of $\delta=0$, the MMDP is transition-independent. We developed a local search algorithm that runs in polynomial time in the number of agents. We derived optimality bounds on the policies obtained from our algorithm as a function of  $\delta$. Our results were verified through numerical studies on a patrolling example and a multi-robot control scenario.

\bibliographystyle{IEEEtran}
\bibliography{cdc2021}

%\newpage
\section*{Appendix}
\emph{Proof of Lemma \ref{lemma:total-variation-bound}:} The total variation distance between these distributions is given by $$\max_{T}{|\sum_{s^{\prime} \in T}{P(s,\pi(s),s^{\prime})-\overline{P}(s,\pi(s),s^{\prime})}|}.$$ With slight abuse of notation, we let $P$ and $\overline{P}$ denote the probability distributions of $s_{t+1}$ when the agents follow policy $\pi$ in MDPs $\mathcal{M}$ and $\hat{\mathcal{M}}$. We let $P(s^{\prime}|s,a)$ (resp. $\overline{P}(s^{\prime}|s,a)$) denote the probability that $s_{t+1} = s^{\prime}$ when $s_{t} = s$ and $a_{t} = a$ in MDP $\mathcal{M}$ (resp. $\hat{\mathcal{M}}$). For a state $s^{\prime} \in S$, we let $s_{1:(i-1)}^{\prime} = \{s_{1}^{\prime},\ldots,s_{i-1}^{\prime}\}$. 

For any $T$, we  define the sets $T_{1:(i-1)}$ for $i=1,\ldots,m$ to denote the set of tuples of $s_{1}^{\prime},\ldots,s_{i-1}^{\prime}$ that can be completed to an element of $T$. We define $T_{i}(s_{1:(i-1)})$ by  $$T_{i}(s_{1:(i-1)}) = \{s_{i}^{\prime} \in S_{i}: \{s_{1:(i-1)}^{\prime},s_{i}^{\prime}\} \subseteq Q \in T\},$$ i.e., $\{s_{1:(i-1)}^{\prime},\ldots,s_{i}^{\prime}\}$ can be completed to an element of $T$. We can then write the probability $P(s^{\prime} \in T | s,\pi(s))$ as $$\prod_{i=1}^{m}{P(s_{i}^{\prime} \in T_{i}(s_{1:(i-1)})|s_{1:(i-1)} \in T_{1:(i-1)},s,\pi(s))}.$$ Hence, the total variation distance is equivalent to 
\begin{IEEEeqnarray*}{rCl}
\IEEEeqnarraymulticol{3}{l}{
\max_{T}{\big|\prod_{i=1}^{m}{P(s_{i}^{\prime}\in T_{i}(s_{1:(i-1)})|s_{1:(i-1)}^{\prime} \in T_{1:(i-1)},s,a)}}} \\
&& - \prod_{i=1}^{m}{\overline{P}(s_{i}^{\prime} \in T_{i}(s_{1:(i-1)}^{\prime})|s_{1:(i-1)}^{\prime} \in T_{1:(i-1)},s,a)}\big| \\
&\leq& \max_{T}{\sum_{i=1}^{m}{|P(s_{i}^{\prime}\in T_{i}(s_{1:(i-1)}|s_{1:(i-1)} \in T_{1:(i-1)})}} \\
&& - \overline{P}(s_{i}^{\prime} \in T_{i}(s_{1:(i-1)}^{\prime})|s_{1:(i-1)}^{\prime} \in T_{1:(i-1)})|
\end{IEEEeqnarray*}
The left-hand side can then be bounded above by
\begin{multline*}%{rCl}
%\IEEEeqnarraymulticol{3}{l}{
    \sum_{i=1}^{m}{\max_{T,U}{|P(s_{i}^{\prime} \in T | s_{-i}^{\prime} \in U, s,a) }}\\
    -\overline{P}(s_{i}^{\prime} \in T | s_{-i}^{\prime} \in U, s,a)|
%    &=& \sum_{i=1}^{m}{\max_{T,U}{|\sum_{s_{-i}^{\prime} \in U}{
\end{multline*}
We have that $\overline{P}$ is transition independent, and moreover 
%{\color{red}
\begin{equation*}
    \overline{P}(s_{i}^{\prime}|s_{-i}^{\prime} \in U, s,a) 
    = P(s_{i}^{\prime}|s_{i},\overline{s}_{-i},\overline{s}_{-i}^{\prime},a_{i},\overline{a}_{-i}).
\end{equation*}
We therefore have 
\begin{multline*}
    \sum_{i=1}^{m}{\max_{T,U}{\big| \sum_{s_{i}^{\prime} \in T}{\left[P(s_{i}^{\prime}|s_{-i}^{\prime} \in U, s, a) \right.}}} \\
    - \left. P(s_{i}^{\prime} | s_{i}, \overline{s}_{-i}, \overline{s}_{-i}^{\prime},a_{i},\overline{a}_{-i}) \right] \big|
\end{multline*}
which is equal to 
\begin{multline*}
    \sum_{i=1}^{m}{\max_{T,U}{\big|\sum_{s_{i}^{\prime} \in T}{\left[\sum_{s_{-i}^{\prime} \in U}{\left(P(s_{i}^{\prime} | s_{-i}^{\prime},s,a)\right)}\right.}}} \\
   \cdot(P(s_{-i}^{\prime}|s_{-i}^{\prime} \in U, s,a)) -P(s_{i}^{\prime}|s_{i},\overline{s}_{-i},\overline{s}_{-i}^{\prime},a_{i},\overline{a}_{-i})]\big|
\end{multline*}
We can then rearrange the order of summation to obtain
\begin{multline*}
    \sum_{i=1}^{m}{\max_{T,U}{\big| \sum_{s_{-i}^{\prime} \in U}{\left[P(s_{-i}^{\prime}|s_{-i}^{\prime} \in U, s, a)\right.}}} \\
    \cdot \left. \sum_{s_{i}^{\prime} \in T}{P(s_{i}^{\prime}|s_{-i}^{\prime},s,a)-P(s_{i}^{\prime}|s_{i},\overline{s}_{-i},\overline{s}_{-i}^{\prime},a_{i},\overline{a}_{-i})}\right] \big|
\end{multline*}
This summation can be bounded above by 
%\begin{multline*}
 %   \sum_{i=1}^{m}{\max_{T,U}{\max_{s_{-i}^{\prime} \in U}{\big| \sum_{s_{i}^{\prime} \in T}{P(s_{i}^{\prime} | s_{-i}^{\prime},s,a)-P(s_{i}^{\prime} | s_{i},\overline{s}_{-i},a_{i},\overline{a}_{-i}, \overline{s}_{-i}^{\prime})} \big|}}}
%\end{multline*}
%and can then be further bounded above by 
\begin{multline*}
    \sum_{i=1}^{m}{\max_{s_{-i}^{\prime},\overline{s}_{-i}^{\prime}}{\max_{T}{\big| \sum_{s_{i}^{\prime} \in T}{(P(s_{i}^{\prime} | s_{-i}^{\prime},s,a) }}}} \\
    -P(s_{i}^{\prime} | s_{i},\overline{s}_{-i},a_{i},\overline{a}_{-i}, \overline{s}_{-i}^{\prime})) \big|
\end{multline*}
The inner maximum is equal to the total variation distance, which is bounded above by $\delta$ by the definition of $\delta$-dependent property.

\end{document}